# Dynamic Evolution Equations for Isolated Smoke Vortexes in Rational Mechanics

Xiao Jianhua

*Measurement Institute, Henan Polytechnic University, Jiaozuo, 454000, China*

**Abstract:** Smoke circle vortexes are a typical dynamic phenomenon in nature. The similar circle vortexes phenomenon appears in hurricane, turbulence, and many others. A semi-empirical method is constructed to get some intrinsic understanding about such circle vortex structures. Firstly, the geometrical motion equations for smoke circle is formulated based empirical observations. Based on them, the mechanic dynamic motion equations are established. Finally, the general dynamic evolution equations for smoke vortex are formulated. They are dynamic evolution equations for exact stress field and dynamic evolution equations for average stress field. For industrial application and experimental data processing, their corresponding approximation equations for viscous fluid are given. Some simple discussions are made.

## Contents

## 1. Introduction

Smoke circle vortexes are a typical dynamic phenomenon in nature. The similar circle vortexes phenomenon appears in hurricane, turbulence, and many others.

Although it is very difficult to get a theoretic solution for smoke circle vortex based on traditional mechanics dynamic equations, however, a semi-empirical method can be constructed to get some intrinsic understanding about such circle vortex structures. This is the main topic for this paper.

Firstly, the geometrical motion equations for smoke circle is formulated based empirical observations. Then, its geometrical structure is exposed. Based on them, the mechanic dynamic motion equations are established. By combining both, the general dynamic evolution equation for smoke vortex is formulated.

As the basic theory has been formulated in other papers by author [1-5], this paper will directly use related results without details.

## 2. Geometrical Structure of Smoke Vortex

For a laminar flow in z direction, a global cylindrical coordinator system ($R, \Theta, Z$) can be established. For simplicity, the Z axe is positioned at the center of smoke vortex. Using the R to represent the center circle radium, a local coordinator system can be established referring to the center circle ($r, \theta, \Theta$). Their relationship is shown in Figure.1 (Coordinator System Definition).
For the local coordinator system, it is embedded on the smoke vortex for instant and so is renewed for the next instant.

Based on empirical observation, the smoke vortex internal motion is described by the deformation tensor defined in local cylindrical coordinator system as:

$$\vec{g}_r = \frac{1}{\cos \varphi}(R_r^r \vec{g}_r^0 + R_r^\theta \vec{g}_\theta^0 + R_r^\Theta \vec{g}_\Theta^0) \tag{1-1}$$

$$\vec{g}_\theta = \frac{1}{\cos \varphi}(R_\theta^r \vec{g}_r^0 + R_\theta^\theta \vec{g}_\theta^0 + R_\theta^\Theta \vec{g}_\Theta^0) \tag{1-2}$$

$$\vec{g}_\Theta = \frac{1}{\cos \varphi}(R_\Theta^r \vec{g}_r^0 + R_\Theta^\theta \vec{g}_\theta^0 + R_\Theta^\Theta \vec{g}_\Theta^0) \tag{1-3}$$

Where, $\vec{g}_r^0$, $\vec{g}_\theta^0$, $\vec{g}_\Theta^0$ are the covariant basic vectors for the local cylindrical coordinator system. After unit time, they are deformed as the instant covariant basic vectors $\vec{g}_r$, $\vec{g}_\theta$, $\vec{g}_\Theta$.

Taking the convention, $dx^3 \to d\Theta$, $dx^2 \to d\theta$, $dx^1 \to dr$, the deformation tensor is expressed as:

$$\frac{1}{\cos\varphi} R^i_j = \frac{1}{\cos\varphi} \begin{vmatrix} \cos\varphi & \sin\varphi & 0 \\ -\sin\varphi & \cos\varphi & 0 \\ 0 & 0 & 1 \end{vmatrix} \quad (2)$$

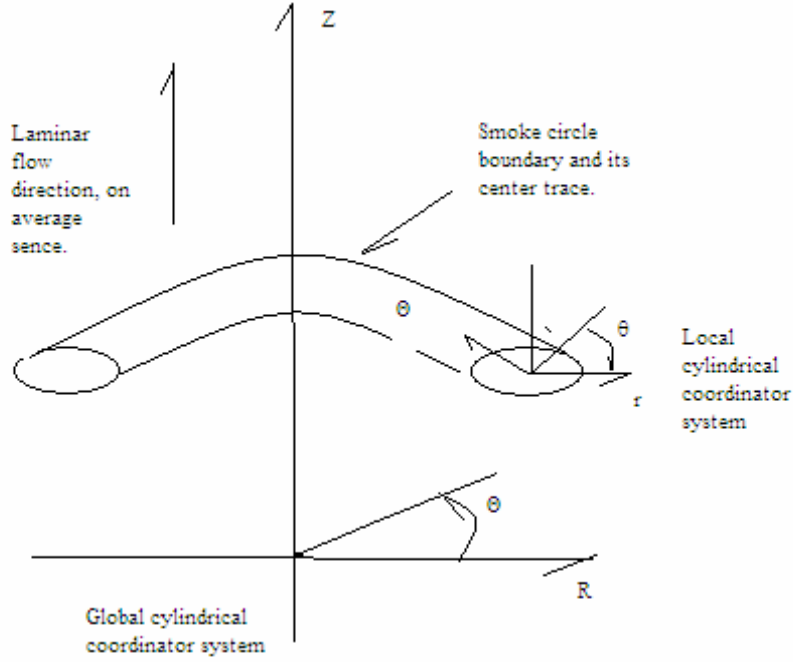

Figure.1 Coordinator System Definition

This is an local rotation with expansion. Based on the related researches done by the author [1], the related velocity fields $u^i$ meet the following equations:

$$u^r\big|_r = 0, \ u^\theta\big|_\theta = 0, \ u^\Theta\big|_\Theta - (\frac{1}{\cos\varphi} - 1) = 0 \quad (3\text{-}1)$$

$$u^r\big|_\theta + u^\theta\big|_r = 0, \ u^r\big|_\Theta \pm u^\Theta\big|_r = 0, \ u^\Theta\big|_\theta \pm u^\theta\big|_\Theta = 0 \quad (3\text{-}2)$$

$$(\frac{1}{\cos\varphi})^2 = 1 + \frac{1}{4}(u^r\big|_\theta - u^\theta\big|_r)^2 \quad (3\text{-}3)$$

Where, $u^i\big|_j$ represents covariant derivatives performed in the local cylindrical coordinator system (initial configuration, or reference configuration for instant deformation). That means the smoke vortex is symmetrical about $\Theta$ coordinator direction.

Note that the local cylindrical coordinator is defined on the cross section of smoke vortex, for the smoke vortex center circle with radium R, one has:

$$g^0_{rr} = 1, \ g^0_{\theta\theta} = r^2, \ g^0_{\Theta\Theta} = R^2 + r^2 + 2rR\cdot\cos\theta \quad (4)$$

Hence, one has:

$$g_0^{rr} = 1, \quad g_0^{\theta\theta} = \frac{1}{r^2}, \quad g_0^{\Theta\Theta} = \frac{1}{R^2 + r^2 + 2rR \cdot \cos\theta} \tag{5}$$

The Christoffel coefficient is obtained as:

$$\Gamma^r_{\theta\theta} = -r, \quad \Gamma^\theta_{r\theta} = \Gamma^\theta_{\theta r} = \frac{1}{r} \tag{6-1}$$

$$\Gamma^r_{\Theta\Theta} = -(r + R \cdot \cos\theta), \quad \Gamma^\Theta_{\Theta r} = \Gamma^\Theta_{r\Theta} = \frac{r + R \cdot \cos\theta}{R^2 + r^2 + 2rR \cdot \cos\theta} \tag{6-2}$$

$$\Gamma^\theta_{\Theta\Theta} = \frac{R}{r}\sin\theta, \quad \Gamma^\Theta_{\Theta\theta} = \Gamma^\Theta_{\theta\Theta} = -\frac{rR \cdot \sin\theta}{R^2 + r^2 + 2rR \cdot \cos\theta} \tag{6-3}$$

The general form of covariant derivative is in form:

$$u^i\big|_j = \frac{\partial u^i}{\partial x^j} + u^k \Gamma^i_{jk} \tag{6-4}$$

The related covariant derivatives are:

$$u^r\big|_r = \frac{\partial u^r}{\partial r} \tag{7-1}$$

$$u^r\big|_\theta = \frac{\partial u^r}{\partial \theta} - r \cdot u^\theta \tag{7-2}$$

$$u^r\big|_\Theta = \frac{\partial u^r}{\partial \Theta} - (r + R \cdot \cos\theta) \cdot u^\Theta \tag{7-3}$$

$$u^\theta\big|_r = \frac{\partial u^\theta}{\partial r} + \frac{1}{r} \cdot u^\theta \tag{7-4}$$

$$u^\theta\big|_\theta = \frac{\partial u^\theta}{\partial \theta} + \frac{1}{r} \cdot u^r \tag{7-5}$$

$$u^\theta\big|_\Theta = \frac{\partial u^\theta}{\partial \Theta} + \frac{R}{r}\sin\theta \cdot u^\Theta \tag{7-6}$$

$$u^\Theta\big|_r = \frac{\partial u^\Theta}{\partial r} + \frac{r + R \cdot \cos\theta}{R^2 + r^2 + 2rR \cdot \cos\theta} \cdot u^\Theta \tag{7-7}$$

$$u^\Theta\big|_\theta = \frac{\partial u^\Theta}{\partial \theta} - \frac{rR \cdot \sin\theta}{R^2 + r^2 + 2rR \cdot \cos\theta} \cdot u^\Theta \tag{7-8}$$

$$u^\Theta\big|_\Theta = \frac{\partial u^\Theta}{\partial \Theta} + \frac{r + R \cdot \cos\theta}{R^2 + r^2 + 2rR \cdot \cos\theta} \cdot u^r - \frac{rR \cdot \sin\theta}{R^2 + r^2 + 2rR \cdot \cos\theta} \cdot u^\theta \tag{7-9}$$

Note that the velocity field is defined by the coordinator measurements not the physical measurements. For example, $u^\theta = \theta(t+1.p) - \theta(t,p)$ for material point $p$, its $\theta$ coordinator is varied after unit time.

(Note: In conventional mechanics theory, the $\tilde{u}^\theta$ is the real physical velocity decomposed component on the $\theta$ coordinator direction. After this convention, the replacement as $\frac{\partial}{\partial \theta} \to \frac{\partial}{r \partial \theta}$, $r u^\theta \to \tilde{u}^\theta$, and $u^r\big|_\theta \to \frac{1}{r} \tilde{u}^r\big|_\theta$, $r u^\theta\big|_r \to \tilde{u}^\theta\big|_r$ are required. Then:

$$u^r\big|_\theta = \frac{\partial u^r}{\partial \theta} - r \cdot u^\theta \to \tilde{u}^r\big|_\theta = \frac{\partial \tilde{u}^r}{r \partial \theta} - \frac{\tilde{u}^\theta}{r}$$

$$u^\theta\Big|_r = \frac{\partial u^\theta}{\partial r} + \frac{1}{r}\cdot u^\theta \rightarrow \tilde{u}^\theta\Big|_r = r(\frac{\partial \frac{\tilde{u}^\theta}{r}}{\partial r} + \frac{1}{r}\cdot\frac{\tilde{u}^\theta}{r}) = \frac{\partial \tilde{u}^\theta}{\partial r}$$

These results are very familiar for mechanics researchers.

Based on tensor theory, there is no need to use physical components. As using the tensor components can simplify the related calculation, the final results be transformed into physical components if there is such requirement. Unfortunately, the apparent difference as the above shows discouraged the wide application of tensor equations in mechanical engineering. It is hopping this paper will promote the application of modern mathematic formulation in mechanical engineering.)

Based on Equations (3), one has the geometrical equations for smoke vortex:

$$\frac{\partial u^r}{\partial r} = 0 \tag{8-1}$$

$$\frac{\partial u^r}{\partial \Theta} - (r + R\cdot\cos\theta)\cdot u^\Theta = 0 \tag{8-2}$$

$$\frac{\partial u^r}{\partial \theta} + \frac{\partial u^\theta}{\partial r} + (\frac{1}{r} - r)\cdot u^\theta = 0 \tag{8-3}$$

$$\frac{\partial u^\theta}{\partial \theta} + \frac{1}{r}\cdot u^r = 0 \tag{8-4}$$

$$\frac{\partial u^\theta}{\partial \Theta} + \frac{R}{r}\sin\theta \cdot u^\Theta = 0 \tag{8-5}$$

$$\left(\frac{1}{\cos\varphi}\right)^2 = 1 + \frac{1}{4}\left(\frac{\partial u^r}{\partial \theta} - \frac{\partial u^\theta}{\partial r} - (\frac{1}{r}+r)\cdot u^\theta\right)^2 \tag{8-6}$$

$$\frac{\partial u^\Theta}{\partial \Theta} + \frac{r+R\cdot\cos\theta}{R^2+r^2+2rR\cdot\cos\theta}\cdot u^r - \frac{rR\cdot\sin\theta}{R^2+r^2+2rR\cdot\cos\theta}\cdot u^\theta = \frac{1}{\cos\varphi} - 1 \tag{8-7}$$

$$\frac{\partial u^\Theta}{\partial \theta} - \frac{rR\cdot\sin\theta}{R^2+r^2+2rR\cdot\cos\theta}\cdot u^\Theta = 0 \tag{8-8}$$

$$\frac{\partial u^\Theta}{\partial r} + \frac{r+R\cdot\cos\theta}{R^2+r^2+2rR\cdot\cos\theta}\cdot u^\Theta = 0 \tag{8-9}$$

Mathematically, the required velocity field components can be determined completely. By observation, (8-8) and (8-9) can give out solution $u^\Theta(r,\theta)$. By observation, (8-1), (8-3), (8-4) and (8-6) can give out solution $u^r(r,\theta)$ and $u^\theta(r,\theta)$; Once they are obtained, (8-2) can give out solution $u^\Theta(r,\theta,\Theta)$; By observation, (8-5) can give out solution $u^\theta(r,\theta,\Theta)$; Then, the equation (8-7) can give out solution $u^\Theta(r,\theta,\Theta)$. All the solutions will take the $R$ and $\varphi$ as their parameters. The final solutions have the form: $u^r(\theta,\Theta)$, $u^\theta(r,\theta,\Theta)$, and $u^\Theta(r,\theta,\Theta)$.

**2.1 Geometrical Equations for Smoke Circle Vortex**

For Equation (8-3) taking partial derivative about $r$, and combining with Equations (8-1), we obtained the equation:

$$\frac{\partial^2 u^\theta}{\partial r^2} + (\frac{1}{r} - r)\frac{\partial u^\theta}{\partial r} - (\frac{1}{r^2} + 1)u^\theta = 0 \tag{9-1}$$

For Equation (8-4) taking partial derivative about $\theta$, and combining with Equations (8-3), we obtained the equation:

$$\frac{\partial^2 u^\theta}{\partial \theta^2} - \frac{1}{r}\frac{\partial u^\theta}{\partial r} + (\frac{1}{r} - r)u^\theta = 0 \tag{9-2}$$

Once $u^\theta = u^\theta(r,\theta,\Theta)$ is obtained, the $u^r = u^r(\theta,\Theta)$ (by Equation (8-1) it is independent about $r$) is obtained by Equation (8-4) as:

$$u^r(\theta,\Theta) = -r \cdot \frac{\partial u^\theta}{\partial \theta} \tag{9-3}$$

Inserting the Equation (8-3) into Equation (8-6), the coupling equation is:

$$\left(\frac{1}{\cos\varphi}\right)^2 = 1 + \left(\frac{\partial u^\theta}{\partial r} + \frac{u^\theta}{r}\right)^2 \tag{9-4}$$

Taking the $\Theta$ as the implied variables, the $u^\theta = u^\theta(r,\theta,\Theta)$, $u^r = u^r(\theta,\Theta)$, and $\frac{1}{\cos\varphi} = \frac{1}{\cos\varphi(r,\theta,\Theta)}$ are obtained. So, Equations (9) form the local rotation dependence on $(r,\theta)$. It is similar with plane deformation problem.

By observation, (8-8) and (8-9) can give out solution $u^\Theta(r,\theta)$:

$$u^\Theta(r,\theta,\Theta) = \frac{\tilde{u}^\Theta(\Theta)}{\sqrt{R^2 + r^2 + 2rR\cos\theta}} \tag{10-1}$$

For Equation (8-7) taking partial derivative about $\Theta$, and combining with Equations (8-2) and (8-5), we obtained the equation:

$$\frac{1}{\sqrt{R^2 + r^2 + 2rR\cos\theta}}\frac{d^2 \tilde{u}^\Theta}{d\Theta^2} + \tilde{u}^\Theta = \frac{\partial}{\partial \Theta}\left(\frac{1}{\cos\varphi}\right) \tag{10-2}$$

It is clear that the global solution is controlled by $\frac{1}{\cos\varphi} = \frac{1}{\cos\varphi(r,\theta,\Theta)}$. The Equations (10) form the local rotation dependence on ($\Theta$).

### 2.2 General Solutions of Static Smoke Circle Vortex

As a special case, the static smoke circle vortex can be defined as:

$$\frac{\partial}{\partial \Theta}(\frac{1}{\cos\varphi(r,\theta,\Theta)}) = 0 \tag{11-1}$$

. It means that: $\varphi = \varphi(r,\theta)$.

The boundary condition is:

$$u^r(\theta,\Theta) = 0, \quad u^\theta(r,\theta,\Theta) = 0, \text{ for } r = 0 \tag{11-2}$$

And:

$$\frac{\partial u^\Theta}{\partial \Theta} = \frac{1}{\cos\varphi_0} - 1, \text{ for } \Theta = 0, \ r = 0 \tag{11-3}$$

This means that the $\Theta = 0$ direction is selected as the same with the physical flow direction of laminar flow field.

The solution will be divided into two parts: 1) motion field on center circle; 2) motion on cross-section

### 2.2.1 Velocity Field along Smoke Circle Center Direction

By Equation (10-2), the macro circulation angle velocity field meets equation:

$$\frac{1}{\sqrt{R^2 + r^2 + 2rR\cos\theta}} \frac{d^2 \tilde{u}^\Theta}{d\Theta^2} + \tilde{u}^\Theta = 0 \tag{12}$$

Its general solution is:

$$\tilde{u}^\Theta(\Theta) = U^\Theta \cdot \exp\left(\sqrt{-1} \cdot \sqrt[4]{R^2 + r^2 + 2rR\cos\theta} \cdot \Theta\right) \tag{13}$$

Where, $U^\Theta(r,\theta)$ is an arbitral function waiting to be defined by initial and boundary conditions. Hence, by Equation (10-1), one has:

$$u^\Theta(r,\theta,\Theta) = U^\Theta \frac{\exp\left(\sqrt{-1} \cdot \sqrt[4]{R^2 + r^2 + 2rR\cos\theta} \cdot \Theta\right)}{\sqrt{R^2 + r^2 + 2rR\cos\theta}} \tag{14}$$

It is a rotation along $\Theta$ direction with angular frequency $\sqrt[4]{R^2 + r^2 + 2rR\cos\theta}$. For the smoke circle center, the angular frequency is $\sqrt{R}$. It shows that bigger circle rotates faster than small circle. (Or say, it shows that the bigger circle has a larger spatial vibration that the smaller circle.) However, the rotate angular velocity amplitude is proportional with $1/R$. It shows that, for small smoke circle, the angular velocity amplitude tends to infinite. In dynamic researches, this is named as the instability or infinity. In statistic researches, the small circle is viewed as random, while the large circle is viewed as coherent structures.

If the physical displacement velocity is used, as the physical component of it is:

$$V^\Theta = \sqrt{g_{\Theta\Theta}^0} \cdot u^\Theta(r,\theta,\Theta) = U^\Theta \cdot \exp\left(\sqrt{-1} \cdot \sqrt[4]{R^2 + r^2 + 2rR\cos\theta} \cdot \Theta\right) \tag{15}$$

The physical rotation velocity along $\Theta$ direction has constant amplitude $U^\Theta$, while its spatial frequency is circle size dependent. Note that, as $V^\Theta = \sqrt{g_{\Theta\Theta}^0} \cdot u^\Theta(r,\theta,\Theta) = \tilde{u}^\Theta(\Theta)$, the Equations (12) and (13) describes physical velocity field component.

By observation Equation (9-4), when Equation (11) holds, one has:

$$\left(\frac{1}{\cos\varphi_0(r,\theta)}\right)^2 = 1 + \left(\frac{\partial u^\theta}{\partial r} + \frac{u^\theta}{r}\right)^2 \tag{16}$$

The smoke velocity fields are decoupled from the smoke circle rotation. In this case, the smoke velocity fields are expressed as: $u^r(\theta)$ and $u^\theta(r,\theta)$.

Therefore, **the smoke circle and the velocity field on the cross-section of the smoke circle have independent geometrical motion equations. Hence, they can coexist. Their coupling is achieved through initial condition or boundary conditions rather than through control** **equations. This gives out an intrinsic explanation about the universal feature of smoke circle vortex and the similar vortex structure in nature (including turbulence).**

Although they are independent in intrinsic structures, they are coupled together in phenomenon layer. To clear this point, consider the center circle ($r = 0$), where $u^r, u^\theta$ are zeros for symmetry consideration. At the center circle, by Equation (8-7), one has:

$$\frac{\partial u^\Theta}{\partial \Theta} = \frac{1}{\cos\varphi(r,\theta)} - 1, \quad \Theta = 0, \quad r = 0 \tag{17}$$

By Equation (14), at $r = 0$:

$$u^\Theta(0, \theta, \Theta) = U^\Theta \frac{\exp(\sqrt{-1} \cdot \sqrt{R} \cdot \Theta)}{R} \tag{18}$$

Hence, by these two equations, on has:

$$U^\Theta = \sqrt{R} \cdot (\frac{1}{\cos\varphi_0} - 1) \tag{19}$$

By equation (15), it is the physical displacement velocity amplitude along $\Theta$ direction. The angular velocity is obtained as:

$$u^\Theta(r, \theta, \Theta) = (\frac{1}{\cos\varphi_0} - 1) \cdot \sqrt{R} \cdot \frac{\exp(\sqrt{-1} \cdot \sqrt[4]{R^2 + r^2 + 2rR\cos\theta} \cdot \Theta)}{\sqrt{R^2 + r^2 + 2rR\cos\theta}} \tag{20}$$

So, the physical displacement velocity field along $\Theta$ direction is:

$$V^\Theta(r, \theta, \Theta) = (\frac{1}{\cos\varphi_0} - 1) \cdot \sqrt{R} \cdot \exp(\sqrt{-1} \cdot \sqrt[4]{R^2 + r^2 + 2rR\cos\theta} \cdot \Theta) \tag{21}$$

Comparing the angular velocity and physical displacement velocity, it can conclude that: **when the circle scale (radium) is small enough, the angular rotational velocity is very large while the physical displacement velocity tends to be infinitesimal small. Or, say, the apparent small physical displacement velocity can contain extremely fast rotating circle vortexes in small scale.**

In fact, this is the main features of turbulence.

*2.2.2 Velocity Fields on Smoke Circle Cross-section*

Referring to this smoke vortex center circle, the velocity field on the cross-section meets equation:

$$\left(\frac{1}{\cos\varphi(r,\theta)}\right)^2 = 1 + \left(\frac{\partial u^\theta}{\partial r} + \frac{u^\theta}{r}\right)^2 \tag{22}$$

The solution about $u^\theta(r, \theta)$ is:

$$u^\theta(r, \theta) = \pm \frac{r}{2} \cdot \sqrt{\left(\frac{1}{\cos\varphi(r,\theta)}\right)^2 - 1} \tag{23}$$

Its physical displacement velocity component is:

$$V^\theta(r, \theta) = \sqrt{g_{\theta\theta}^0} \cdot u^\theta(r, \theta) = \pm \frac{r^2}{2} \cdot \sqrt{\left(\frac{1}{\cos\varphi}\right)^2 - 1} \tag{24}$$

It shows that the physical displacement velocity is increased from zero at the center circle to maximum at the vortex boundary.

By Equation (9-3), the other velocity component is:

$$u^r(\theta) = -r \cdot \frac{\partial u^\theta}{\partial \theta} = \mp \frac{r^2}{2} \frac{\partial}{\partial \theta} \sqrt{(\frac{1}{\cos\varphi})^2 - 1} \qquad (25)$$

Note that, by Equation (8-1), $u^r(\theta)$ is independent from $r$. So, by the above equation, one has the conclusion: $\frac{\partial}{\partial \theta} \sqrt{(\frac{1}{\cos\varphi})^2 - 1} \propto \frac{1}{r^2}$.

Its physical displacement velocity component is the same as $\sqrt{g_{rr}^0} = 1$. That is:

$$V^r(\theta) = \mp \frac{r^2}{2} \frac{\partial}{\partial \theta} \sqrt{(\frac{1}{\cos\varphi})^2 - 1} \qquad (26)$$

They show that: once the local rotation angle function $\varphi(r,\theta)$ is given, the velocity fields on vortex cross-section are determined.

For special cases, $\frac{\partial}{\partial \theta} \sqrt{(\frac{1}{\cos\varphi})^2 - 1} \approx 0$, that is when the smoke vortex is symmetry about its center circle, one has:

$$V^r(\theta) = 0, \quad u^r(\theta) = 0 \qquad (27)$$

The zero velocity on cross-section radium direction means that the smoke vortex boundary is fixed. The velocity field on the cross-section is a rotation field around the center circle line.

However, for general cases, when the rotation symmetry is broken, the velocity on cross-section radium direction has two kind typical solutions:

### 2.2.3 Smoke Vortex Forming Conditions

The first kind solution is smoke vortex forming solution:

For $u^\theta > 0$, if $\frac{\partial}{\partial \theta} \sqrt{(\frac{1}{\cos\varphi})^2 - 1} > 0$, then $u^r < 0$. For $u^\theta < 0$, if $\frac{\partial}{\partial \theta} \sqrt{(\frac{1}{\cos\varphi})^2 - 1} < 0$, then $u^r < 0$.

The smoke vortex is contracting into center circle position. As a result, the smoke vortex will be more and more significant. It corresponds to the smoke circle vortex forming process.

Hence, the smoke circle vortex forming conditions are:

$$\frac{\partial u^\Theta}{\partial \Theta} = \frac{1}{\cos\varphi} - 1 > 0, \quad u^\theta > 0, \quad \frac{\partial}{\partial \theta} \sqrt{(\frac{1}{\cos\varphi})^2 - 1} > 0, \quad u^r < 0 \qquad (28\text{-}1)$$

$$\frac{\partial u^\Theta}{\partial \Theta} = \frac{1}{\cos\varphi} - 1 > 0, \quad u^\theta < 0, \quad \frac{\partial}{\partial \theta} \sqrt{(\frac{1}{\cos\varphi})^2 - 1} < 0, \quad u^r < 0 \qquad (28\text{-}2)$$

Referring to the local rotation spatial acceleration along $\Theta$ direction, taking the coordinator right-hand chirality selection, the smoke circle vortex forming conditions are: on the cross-section, around the center circle, the flow rotation direction and the local rotation increase direction are the same.

### *2.2.4 Smoke Vortex Broken Conditions*

The second solution is smoke vortex broken process solution:

For $u^\theta > 0$, if $\frac{\partial}{\partial \theta}\sqrt{(\frac{1}{\cos\varphi})^2 - 1} < 0$, then $u^r > 0$. For $u^\theta < 0$, if $\frac{\partial}{\partial \theta}\sqrt{(\frac{1}{\cos\varphi})^2 - 1} > 0$, then $u^r > 0$.

The smoke vortex is expanded outward from the center circle position. As a result, the smoke vortex will be less and less significant. It corresponds to the smoke circle vortex disappearing process.

Hence, the smoke circle vortex broken conditions are:

$$\frac{\partial u^\Theta}{\partial \Theta} = \frac{1}{\cos\varphi} - 1 > 0, \ u^\theta > 0, \ \frac{\partial}{\partial \theta}\sqrt{(\frac{1}{\cos\varphi})^2 - 1} < 0, \ u^r > 0 \qquad (29\text{-}1)$$

$$\frac{\partial u^\Theta}{\partial \Theta} = \frac{1}{\cos\varphi} - 1 > 0, \ u^\theta < 0, \ \frac{\partial}{\partial \theta}\sqrt{(\frac{1}{\cos\varphi})^2 - 1} > 0, \ u^r > 0 \qquad (29\text{-}2)$$

Referring to the local rotation spatial acceleration along $\Theta$ direction, taking the coordinator right-hand chirality selection, the smoke circle vortex forming conditions are: on the cross-section, around the center circle, the flow rotation direction and the local rotation increase direction are the inverse.

### *2.2.5 Smoke Vortex Maintaining Conditions*

Once a smoke vortex is formed, its maintaining condition is:

$$\frac{\partial}{\partial \theta}\sqrt{(\frac{1}{\cos\varphi})^2 - 1} = 0 \qquad (30)$$

Summary above results, when there is a spatial acceleration of velocity, (say $\frac{\partial u^\Theta}{\partial \Theta} = \frac{1}{\cos\varphi} - 1 > 0$), on a plane taking this direction as normal (say $(r,\theta)$), when the velocity fields meet some conditions, smoke circle vortex may be formed, be broken or maintained which depends on the velocity components direction and the spatial acceleration of velocity on the plane.

The natural question is that: for arbitral flow field, how to formulate the smoke vortex related quantity $\frac{1}{\cos\varphi}$. This problem is answered in the next subsection.

### **2.3 General Geometrical Solutions for Smoke Circle Vortex**

Observing the solution for Velocity Fields on Smoke Circle Cross-section, once $\frac{1}{\cos\varphi}$ is given, their solutions are determined. Hence, by insertion them into the general equation (Equation (8-7)):

$$\frac{\partial u^\Theta}{\partial \Theta} + \frac{r + R\cdot\cos\theta}{R^2 + r^2 + 2rR\cdot\cos\theta}\cdot u^r - \frac{rR\cdot\sin\theta}{R^2 + r^2 + 2rR\cdot\cos\theta}\cdot u^\theta = \frac{1}{\cos\varphi} - 1 \qquad (31)$$

One has:

$$\frac{\partial u^{\Theta}}{\partial \Theta} \mp \frac{r^2(r+R\cdot\cos\theta)}{2(R^2+r^2+2rR\cdot\cos\theta)} \frac{\partial}{\partial \theta}\sqrt{(\frac{1}{\cos\varphi})^2-1} \mp \frac{r^2 R\cdot\sin\theta}{2(R^2+r^2+2rR\cdot\cos\theta)}\cdot\sqrt{(\frac{1}{\cos\varphi})^2-1} = \frac{1}{\cos\varphi}-1$$

(32)

This equation will give out general function solutions:

$$T[u^{\Theta}(r,\theta,\Theta), \frac{1}{\cos\varphi(r,\theta,\Theta)}] = 0 \tag{33}$$

Taking the velocity field $u^{\Theta}$ as given field, this general function can be used to evaluate the smoke vortex evolution conditions.

*2.3.1 Determine Cross-section Solution by Known Circle Direction Velocity Field*

For the given velocity field $u^{\Theta}$, the $\frac{1}{\cos\varphi}$ can be obtained, through the general function solutions: $T[u^{\Theta}(r,\theta,\Theta), \frac{1}{\cos\varphi(r,\theta,\Theta)}] = 0$. Hence, the smoke vortex solution is obtained by above two equations.

$$u^{\theta}(r,\theta) = \pm\frac{r}{2}\cdot\sqrt{\left(\frac{1}{\cos\varphi}\right)^2-1} \tag{34}$$

$$u^r(\theta) = \mp\frac{r^2}{2}\frac{\partial}{\partial\theta}\sqrt{(\frac{1}{\cos\varphi})^2-1} \tag{35}$$

*2.3.2 Determine Circle Direction Velocity Field by Known Cross-section Fields*

If the velocity field $u^{\theta}$ and $u^r$ are given fields, then the $\frac{1}{\cos\varphi}$ is obtained through following equations.

$$\left(\frac{1}{\cos\varphi}\right)^2 = 1 + \frac{1}{4}\left(\frac{\partial u^r}{\partial\theta} - \frac{\partial u^{\theta}}{\partial r} - (\frac{1}{r}+r)\cdot u^{\theta}\right)^2 \tag{36-1}$$

Or,

$$\left(\frac{1}{\cos\varphi}\right)^2 = 1 + \left(\frac{\partial u^{\theta}}{\partial r} + \frac{u^{\theta}}{r}\right)^2 \tag{36-2}$$

By the obtained $\frac{1}{\cos\varphi}$, through the general function solutions: $T[u^{\Theta}(r,\theta,\Theta), \frac{1}{\cos\varphi(r,\theta,\Theta)}] = 0$, the circle direction velocity field is obtained.

In fact, the center circle solution is:

$$u^{\Theta}(r,\theta,\Theta) = (\frac{1}{\cos\varphi_0}-1)\cdot\sqrt{R}\cdot\frac{\exp\left(\sqrt{-1}\cdot\sqrt[4]{R^2+r^2+2rR\cos\theta}\cdot\Theta\right)}{\sqrt{R^2+r^2+2rR\cos\theta}} \tag{37}$$

*2.3.3 Multiple-scale Solutions for Cascade Smoke Vortexes*

Based on above research, any spatial acceleration may produce smoke vortexes. Therefore, when $\frac{\partial u^r}{\partial r}$ or $\frac{\partial u^{\theta}}{\partial \theta}$ is not zero, sub-scale smoke vortexes may be generated. The above methods can be used again (with coordinator system modification) to obtain the sub=scale smoke vortexes solution.

Repeating this process, multiple-scale solutions for cascade smoke vortexes are obtained.

This explains the universal existence of cascade vortexes phenomenon.

**2.4 Features on Laboratory Measure of Smoke Circle Displacement Velocity Fields**

For the coordinator system ($R, \Theta, Z$), taking the X coordinator along R direction at $\Theta = 0$, then a right-hand chirality rectangular laboratory coordinator system is established as ($X, Y, Z$). In this laboratory system, the physical field $V^\Theta$ is observed as:

$$V_X^\Theta = (\frac{1}{\cos\varphi_0} - 1) \cdot \sqrt{R} \cdot \cos\left(\sqrt[4]{R^2 + r^2 + 2rR\cos\theta} \cdot \Theta\right) \tag{38-1}$$

$$V_y^\Theta = (\frac{1}{\cos\varphi_0} - 1) \cdot \sqrt{R} \cdot \sin\left(\sqrt[4]{R^2 + r^2 + 2rR\cos\theta} \cdot \Theta\right) \tag{38-2}$$

Taking the $r$ direction parallel with the R direction, using the right-hand chirality, the physical displacement velocity field $V^r, V^\theta$ can also be expressed in laboratory coordinator system as:

$$V_X^r = \mp \frac{r^2}{2} \frac{\partial}{\partial \theta} \sqrt{(\frac{1}{\cos\varphi})^2 - 1} \cdot \cos\theta \cdot \cos\Theta \tag{38-3}$$

$$V_Y^r = \mp \frac{r^2}{2} \frac{\partial}{\partial \theta} \sqrt{(\frac{1}{\cos\varphi})^2 - 1} \cdot \cos\theta \cdot \sin\Theta \tag{38-4}$$

$$V_Z^r = \pm \frac{r^2}{2} \frac{\partial}{\partial \theta} \sqrt{(\frac{1}{\cos\varphi})^2 - 1} \cdot \sin\theta \tag{38-5}$$

$$V_X^\theta = \mp \frac{r^2}{2} \cdot \sqrt{\left(\frac{1}{\cos\varphi}\right)^2 - 1} \cdot \sin\theta \cdot \cos\Theta \tag{38-6}$$

$$V_Y^\theta = \mp \frac{r^2}{2} \cdot \sqrt{\left(\frac{1}{\cos\varphi}\right)^2 - 1} \cdot \sin\theta \cdot \sin\Theta \tag{38-7}$$

$$V_Z^\theta = \pm \frac{r^2}{2} \cdot \sqrt{\left(\frac{1}{\cos\varphi}\right)^2 - 1} \cdot \cos\theta \tag{38-8}$$

Hence, the total physical displacement velocity fields observed in laboratory will be very complicated as:

$$V_X = V_X^r + V_X^\theta + V_X^\Theta \tag{39-1}$$

$$V_Y = V_Y^r + V_Y^\theta + V_Y^\Theta \tag{39-2}$$

$$V_Z = V_X^r + V_X^\theta + V_Z^0 \tag{39-3}$$

Where, the $V_Z^0$ is the laminar flow physical displacement velocity.

Observing the related items, as a first order approximation, omitting the items related with $\frac{\partial}{\partial \theta} \sqrt{(\frac{1}{\cos \varphi})^2 - 1}$, the approximation solution in laboratory system is:

$$V_X = (\frac{1}{\cos \varphi_0} - 1) \cdot \sqrt{R} \cdot \cos\left(\sqrt[4]{R^2 + r^2 + 2rR\cos\theta} \cdot \Theta\right) \mp \frac{r^2}{2} \cdot \sqrt{\left(\frac{1}{\cos \varphi_0}\right)^2 - 1} \cdot \sin\theta \cdot \cos\Theta \quad (40\text{-}1)$$

$$V_Y = (\frac{1}{\cos \varphi_0} - 1) \cdot \sqrt{R} \cdot \sin\left(\sqrt[4]{R^2 + r^2 + 2rR\cos\theta} \cdot \Theta\right) \mp \frac{r^2}{2} \cdot \sqrt{\left(\frac{1}{\cos \varphi_0}\right)^2 - 1} \cdot \sin\theta \cdot \sin\Theta \quad (40\text{-}2)$$

$$V_Z = V_Z^0 \pm \frac{r^2}{2} \cdot \sqrt{\left(\frac{1}{\cos \varphi_0}\right)^2 - 1} \cdot \cos\theta \quad (40\text{-}3)$$

The large scale structure has the amplitude $(\frac{1}{\cos \varphi_0} - 1) \cdot \sqrt{R}$, while the substructure has the amplitude $\frac{r^2}{2} \cdot \sqrt{\left(\frac{1}{\cos \varphi_0}\right)^2 - 1}$. Their phase relation is very complicated.

    It can be explained as: on macro spatial irregular shear flow, a small regular rotation with spatial dependence is staked. The laminar flow is disturbed by the small regular rotation with spatial variant amplitude. The above solutions are summarized in Figure.2 (Velocity Field of Simple Smoke Vortex).

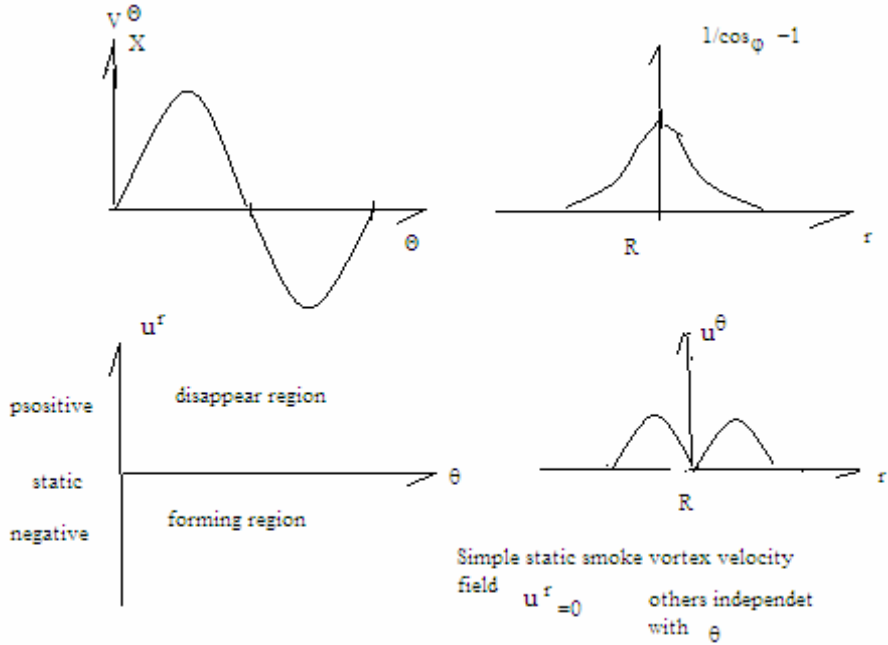

Figure.2 Velocity Field of Simple Smoke Vortex

## 3. Stress Fields for Smoke Circle Vortex

    For the smoke circle vortex, the intrinsic stress is defined by the physical deformation tensor.

Based on deformation Equation (2), the physical deformation expressed by the physical displacement velocity is obtained through operation:

$$\tilde{F}^i_j = \frac{\sqrt{g^0_{(ii)}}}{\sqrt{g^0_{(jj)}}} (\frac{1}{\cos\varphi} R^i_j - \delta^i_j) = \begin{vmatrix} 0 & \frac{1}{r}\tan\varphi & 0 \\ -r\cdot\tan\varphi & 0 & 0 \\ 0 & 0 & \frac{1}{\cos\varphi}-1 \end{vmatrix} \quad (41)$$

Letting the medium parameters are $(\tilde{\lambda}, \tilde{\mu})$ as Newton fluid, the non zero physical stress field components are:

$$\sigma^r_r = \sigma^\theta_\theta = \tilde{\lambda}\left(\frac{1}{\cos\varphi}-1\right) \quad (42\text{-}1)$$

$$\sigma^\Theta_\Theta = (\tilde{\lambda}+2\tilde{\mu})\left(\frac{1}{\cos\varphi}-1\right) \quad (42\text{-}2)$$

$$\sigma^r_\theta = 2\tilde{\mu}\cdot\frac{1}{r}\cdot\tan\varphi \quad (42\text{-}3)$$

$$\sigma^\theta_r = -2\tilde{\mu}\cdot r\cdot\tan\varphi \quad (42\text{-}4)$$

Based on this formulation, some basic conclusions can be obtained.

### 3.1 Stream Circle Line Flow Viewpoint and Zero Pressure Surface

Referring to the initial static pressure: $-p_0\delta^i_j$, view the pressure variation caused by smoke vortex as: $\tilde{\lambda}\left(\frac{1}{\cos\varphi}-1\right)\delta^i_j$, then in the smoke vortex region, the current dynamic pressure is:

$$-p\delta^i_j = -p_0\delta^i_j + \tilde{\lambda}\left(\frac{1}{\cos\varphi}-1\right)\delta^i_j \quad (43\text{-}1)$$

It shows that the effective pressure is largely reduced.

Except this pressure variation, the remaining stress components are:

$$\Delta\sigma^\Theta_\Theta = 2\tilde{\mu}\cdot\left(\frac{1}{\cos\varphi}-1\right) \quad (43\text{-}2)$$

This condition means that there is net force acting on the cross-section along the smoke circle rotation direction.

The zero pressure condition equation is given by a critical local rotation angle $\varphi_c$ defined by equation:

$$\tilde{\lambda}\left(\frac{1}{\cos\varphi_c}-1\right) - p_0 = 0 \quad (44\text{-}1)$$

At the region $\varphi < \varphi_c$, the pressure is negative as normal fluid. At the region $\varphi > \varphi_c$, the pressure is positive as un-normal fluid. The zero pressure surface is defined by surface equation:

$$\varphi(r,\theta,\Theta) = \varphi_c \quad (44\text{-}2)$$

In fact, for industry application, the zero stress surface plays an important role.

### 3.2 Cross-section Plane Flow Viewpoint and Zero Pressure Surface

1). Plane flow view-point. For static initial pressure: $-p_0\delta^i_j$, in the smoke vortex region,

define the current dynamic pressure as:

$$-p\delta_j^i = -p_0 \delta_j^i + (\tilde{\lambda} + 2\tilde{\mu})\left(\frac{1}{\cos\varphi} - 1\right)\delta_j^i \tag{45-1}$$

Except this pressure variation, the remaining stress components are:

$$\Delta\sigma_r^r = \Delta\sigma_\theta^\theta = -2\tilde{\mu}\cdot\left(\frac{1}{\cos\varphi} - 1\right) \tag{45-2}$$

This condition means that there are surface forces acting on the vortex boundary surface. The flow field is on the cross-section plane.

The zero pressure condition equation is given by a critical local rotation angle $\tilde{\varphi}_c$ defined by equation:

$$(\tilde{\lambda} + 2\tilde{\mu})\left(\frac{1}{\cos\tilde{\varphi}_c} - 1\right) - p_0 = 0 \tag{46-1}$$

At the region $\varphi < \tilde{\varphi}_c$, the pressure is negative as normal fluid. At the region $\varphi > \tilde{\varphi}_c$, the pressure is positive as un-normal fluid. The zero pressure surface is defined by surface equation:

$$\varphi(r,\theta,\Theta) = \tilde{\varphi}_c \tag{46-2}$$

This zero pressure is different from the above one. In fact, for industry application, the zero stress surface selection depends on what flow paten is important.

**3.3 Boundary Condition Viewpoint**

Except the normal stress components conditions, the shear stress conditions can be formulated. They are:

$$\sigma_\theta^r = 2\tilde{\mu}\cdot\frac{1}{r}\cdot\tan\varphi = 2\tilde{\mu}\cdot\frac{1}{r}\cdot\frac{\sin\varphi}{\cos\varphi} \tag{47-1}$$

$$\sigma_r^\theta = -2\tilde{\mu}\cdot r\cdot\tan\varphi = -2\tilde{\mu}\cdot r\cdot\frac{\sin\varphi}{\cos\varphi} \tag{47-2}$$

Firstly, the shear stress is not symmetrical. This conclusion has be verified by many experiments.

Introducing the average angular velocity field $\bar{u}^\Theta$, the average physical displacement velocity $\bar{V}^\Theta = R\cdot\bar{u}^\Theta$, defining the spatial wave number of smoke circle vortex as:

$$k = \frac{\frac{\partial u^\Theta}{\partial \Theta}}{\bar{u}^\Theta} = \left(\frac{1}{\cos\varphi} - 1\right)\cdot\frac{R}{\bar{V}^\Theta} \tag{48-1}$$

One has:

$$\cos\varphi = \frac{R}{R + k\cdot\bar{V}^\Theta} \tag{48-2}$$

$$\sin\varphi = \sqrt{1 - \left(\frac{R}{R + k\cdot\bar{V}^\Theta}\right)^2} \tag{48-3}$$

By wave number parameter, the shear stresses are formulated as:

$$\sigma_\theta^r = 2\tilde{\mu}\cdot\frac{1}{r}\cdot\frac{\sin\varphi}{\cos\varphi} = 2\tilde{\mu}\cdot\frac{1}{r}\cdot\sqrt{\left(1 + \frac{k\cdot\bar{V}^\Theta}{R}\right)^2 - 1} \tag{49-1}$$

$$\sigma_r^\theta = -2\tilde{\mu}\cdot r\cdot\frac{\sin\varphi}{\cos\varphi} = -2\tilde{\mu}\cdot r\cdot\sqrt{\left(1 + \frac{k\cdot\bar{V}^\Theta}{R}\right)^2 - 1} \tag{49-2}$$

The characteristic length unit can be defined as:

$$l_I(k) = \sqrt{(1+\frac{k\cdot \overline{V}^\Theta}{R})^2 -1} \qquad (50)$$

For $r \gg l_I(k)$ (a scale much larger than characteristic length unit), the stress boundary condition Equation (47) can be simplified as:

$$\sigma_\theta^r = 0, \quad \sigma_r^\theta = -2\tilde{\mu}\cdot l_I(k)\cdot r, \text{ external boundary} \qquad (51)$$

This condition can be explained as far external boundary shear stress condition. For viscous fluid, the boundary drag force will produce this condition.

For $r \ll l_I(k)$ (a scale much smaller than characteristic length unit), the stress boundary condition can be simplified as:

$$\sigma_\theta^r = 2\tilde{\mu}\cdot l(k)\cdot \frac{1}{r}, \quad \sigma_r^\theta = 0, \text{ internal boundary} \qquad (52)$$

This condition can be explained as internal boundary shear stress condition. When a heavy stone drops into static flow fluid, this stress condition will be meets.

In fact, the traditional description of turbulence production is formulated by the plane shear flow view-point.

For high-speed or highly spatial irregular flow, when $\frac{k\cdot \overline{V}^\Theta}{R} \gg 1$, the characteristic length unit can be approximated as:

$$l_I(k) \approx \frac{k\cdot \overline{V}^\Theta}{R} \qquad (53)$$

This is the high-speed flow characteristic length unit. The turbulence is caused by boundary shear force.

For low-speed or highly spatial regular flow, when $\frac{k\cdot \overline{V}^\Theta}{R} \ll 1$, the characteristic length unit can be approximated as:

$$l_I(k) \approx \sqrt{\frac{2k\cdot \overline{V}^\Theta}{R}} \qquad (54)$$

The turbulence is caused by internal shear force. Little study has been made for this phenomenon.

Note that the mean square shear stress variation in smoke circle vortex can be defined as:

$$\overline{\sigma} = \sqrt{|\sigma_r^\theta \sigma_\theta^r|} = 2\tilde{\mu}\cdot l_I(k) \qquad (55)$$

In this formulation, for very high-speed flow, the usual Reynolds Number is equivalent with:

$$\text{Re} = \frac{\overline{\sigma}}{2\tilde{\mu}k} = \frac{l_I(k)}{k} \approx \frac{\overline{V}^\Theta}{R} \qquad (56)$$

Then, the mechanics meaning of Reynolds Number is the mean square stress variation on boundary surface of smoke vortex. If it is very large, the smoke vortex will be very significant.

Referring the initial background static pressure, another Stress Contrast Number can be defined as:

$$Sc = \frac{\overline{\sigma}}{kp_0} = \frac{2\tilde{\mu}\cdot l_I(k)}{kp_0} \approx \frac{2\tilde{\mu}}{p_0}\cdot \frac{\overline{V}^\Theta}{R} \qquad (57)$$

This is a dimensionless parameter. Its mechanics meaning is the percentage of the mean square stress variation on boundary surface of smoke vortex comparing with the initial static pressure of fluid.

Based on author's viewpoint, the Stress Contrast Number should be used to evaluate the stress variation strength of smoke vortex.

## 4. Dynamic Motion Equations for Smoke Vortex

Here, for logic consistence, the stress defined in dragging coordinator system will be used in motion equations. The physical components will be obtained from the final results.

### 4.1 Notes on Material Invariance

At this point, it is valuable to point out that: the material feature invariance means, for physical constants $(\tilde{\lambda}, \tilde{\mu})$, the physical stress $\sigma_j^i$ is related with the dragging system stress $\tilde{\sigma}_j^i$:

$\sigma_j^i = \frac{\sqrt{g_{(ii)}^0}}{\sqrt{g_{(jj)}^0}} \tilde{\sigma}_j^i$. For isotropic fluid, the constitutive equation is:

$$\sigma_j^i = \frac{\sqrt{g_{(ii)}^0}}{\sqrt{g_{(jj)}^0}} \tilde{\sigma}_j^i = -p_0 \delta_j^i + \tilde{\lambda}(\Delta)\delta_j^i + 2\tilde{\mu} \cdot \frac{\sqrt{g_{(ii)}^0}}{\sqrt{g_{(jj)}^0}} \cdot u^i \Big|_j \tag{58}$$

Where, $\Delta = u^1\big|_1 + u^2\big|_2 + u^3\big|_3 = \frac{\sqrt{g_{11}^0}}{\sqrt{g_{11}^0}} u^1\big|_1 + \frac{\sqrt{g_{22}^0}}{\sqrt{g_{22}^0}} u^2\big|_2 + \frac{\sqrt{g_{33}^0}}{\sqrt{g_{33}^0}} u^3\big|_3$ is invariant about the gauge tensor selection. Hence, it is easy to see that: the Equation (58) is equivalent to equation:

$$\tilde{\sigma}_j^i = -p_0 \delta_j^i + \tilde{\lambda}(\Delta)\delta_j^i + 2\tilde{\mu} \cdot u^i \Big|_j \tag{59}$$

It shows that, for dragging coordinator system, the material features are invariant for the deformation process once the initial coordinator system is selected.

C. Truesdell and W. Noll want to promote this point as the material invariance principle. However, their efforts are mostly omitted.

### 4.2 Dynamic Motion Equations

In dragging coordinator system, for linear moment conservation, the dynamic motion equation is:

$$\tilde{\sigma}_j^i \Big|_j = \rho_0 \frac{\partial u^i}{\partial t} \tag{60-1}$$

For angular moment conservation, the dynamic motion equation is:

$$\tilde{\sigma}_j^i \Big|_i = \rho_0 \frac{\partial}{\partial t}(\frac{1}{\cos\varphi} u^i g_{il}^0 R_j^l) \tag{60-2}$$

For the details for these two equations, please refer [1].

The intrinsic stress defined in dragging coordinator system is:

$$\tilde{\sigma}_r^r = \tilde{\sigma}_\theta^\theta = \tilde{\lambda}\left(\frac{1}{\cos\varphi} - 1\right) \tag{61-1}$$

$$\tilde{\sigma}_\Theta^\Theta = (\tilde{\lambda} + 2\tilde{\mu})\left(\frac{1}{\cos\varphi} - 1\right) \tag{61-2}$$

$$\sigma_\theta^r = 2\tilde{\mu} \cdot \tan\varphi \tag{61-3}$$

$$\sigma_r^\theta = -2\tilde{\mu} \cdot \tan\varphi \tag{61-4}$$

The geometrical deformation equation is:

$$\frac{1}{\cos\varphi} R_j^i = \frac{1}{\cos\varphi} \begin{vmatrix} \cos\varphi & \sin\varphi & 0 \\ -\sin\varphi & \cos\varphi & 0 \\ 0 & 0 & 1 \end{vmatrix} \tag{61-5}$$

$$g_{rr}^0 = 1, \quad g_{\theta\theta}^0 = r^2, \quad g_{\Theta\Theta}^0 = R^2 + r^2 + 2rR\cdot\cos\theta \tag{61-6}$$

$$\frac{1}{\cos\varphi} g_{il}^0 R_j^l = \begin{vmatrix} 1 & 0 & 0 \\ 0 & r^2 & 0 \\ 0 & 0 & \dfrac{R^2 + r^2 + 2rR\cos\theta}{\cos\varphi} \end{vmatrix} \tag{61-7}$$

The related Christoffel coefficient is obtained as:

$$\Gamma_{\theta\theta}^r = -r, \quad \Gamma_{r\theta}^\theta = \Gamma_{\theta r}^\theta = \frac{1}{r} \tag{61-8}$$

$$\Gamma_{\Theta\Theta}^r = -(r + R\cdot\cos\theta), \quad \Gamma_{\Theta r}^\Theta = \Gamma_{r\Theta}^\Theta = \frac{r + R\cdot\cos\theta}{R^2 + r^2 + 2rR\cdot\cos\theta} \tag{61-9}$$

$$\Gamma_{\Theta\Theta}^\theta = \frac{R}{r}\sin\theta, \quad \Gamma_{\Theta\theta}^\Theta = \Gamma_{\theta\Theta}^\Theta = -\frac{rR\cdot\sin\theta}{R^2 + r^2 + 2rR\cdot\cos\theta} \tag{61-10}$$

The covariant derivatives of stress tensor are defined as:

$$\tilde\sigma_j^i\big|_k + \frac{\partial \tilde\sigma_j^i}{\partial x^k} + \tilde\sigma_j^m \Gamma_{km}^i - \tilde\sigma_m^i \Gamma_{jk}^m \tag{61-11}$$

After an algebra operation, the motion equations for linear momentum conservation are obtained as:

$$\frac{\partial \tilde\sigma_r^r}{\partial r} + \frac{\partial \tilde\sigma_\theta^r}{\partial \theta} + (2r + R\cos\theta)\tilde\sigma_r^r - \frac{R}{r}\sin\theta\cdot\tilde\sigma_\theta^r - r\cdot\tilde\sigma_\theta^\theta - (r + R\cos\theta)\tilde\sigma_\Theta^\Theta = \rho_0 \frac{\partial u^r}{\partial t} \tag{62-1}$$

$$\frac{\partial \tilde\sigma_r^\theta}{\partial r} + \frac{\partial \tilde\sigma_\theta^\theta}{\partial \theta} + (2r + R\cos\theta)\tilde\sigma_r^\theta + \frac{R}{r}\sin\theta\cdot(\tilde\sigma_\theta^\theta - \tilde\sigma_\Theta^\Theta) = \rho_0 \frac{\partial u^\theta}{\partial t} \tag{62-2}$$

$$\frac{\partial \tilde\sigma_\Theta^\Theta}{\partial \Theta} = \rho_0 \frac{\partial u^\Theta}{\partial t} \tag{62-3}$$

The motion equations for angular momentum conservation are obtained as:

$$\frac{\partial \tilde\sigma_r^r}{\partial r} + \frac{\partial \tilde\sigma_r^\theta}{\partial \theta} + (\frac{1}{r} + \frac{r + R\cos\theta}{R^2 + r^2 + 2rR\cos\theta})\tilde\sigma_r^r$$
$$-\frac{rR\sin\theta}{R^2 + r^2 + 2rR\cos\theta}\tilde\sigma_r^\theta - \frac{1}{r}\tilde\sigma_\theta^\theta - \frac{r + R\cos\theta}{R^2 + r^2 + 2rR\cos\theta}\tilde\sigma_\Theta^\Theta = \rho_0 \frac{\partial u^r}{\partial t} \tag{63-1}$$

$$\frac{\partial \tilde\sigma_\theta^r}{\partial r} + \frac{\partial \tilde\sigma_\theta^\theta}{\partial \theta} + \frac{r + R\cos\theta}{R^2 + r^2 + 2rR\cos\theta}\tilde\sigma_\theta^r$$
$$-\frac{rR\sin\theta}{R^2 + r^2 + 2rR\cos\theta}\tilde\sigma_\theta^\theta + \frac{rR\sin\theta}{R^2 + r^2 + 2rR\cos\theta}\tilde\sigma_\Theta^\Theta = \rho_0 r^2 \cdot \frac{\partial u^\theta}{\partial t} \tag{63-2}$$

$$\frac{\partial \tilde\sigma_\Theta^\Theta}{\partial \Theta} = \rho_0 \frac{\partial}{\partial t}(\frac{R^2 + r^2 + 2rR\cos\theta}{\cos\varphi} \cdot u^\Theta) \tag{63-3}$$

It is clear, if the Equation (62-3) is satisfied, the Equation (63-3) may not be satisfied. This means that: the deformation is instable as there is dynamic singularity on force balance.

In fact, the Equations (62) are linear momentum conservation about initial reference configuration. The Equations (63) are angular momentum conservation about current reference configuration. They have different physical consideration.

As they may not be met at the same time, the only physical reasons are **the smoke circle vortex is intrinsically instable**. That is to say, **it is a singularity of general motion**.

**4.3 Physical Stability Condition of Smoke Vortex and Evolution Equations**

Based on empirical observation, the physical stable condition is:

$$\frac{\partial}{\partial t}(\frac{R^2+r^2+2rR\cos\theta}{\cos\varphi}\cdot u^\Theta)-\frac{\partial u^\Theta}{\partial t}=0 \tag{64}$$

This means that the velocity $u^\Theta$ is harmonic decaying or increasing about time and space, or simply in harmonic. Furthermore, by Equation (12), $u^\Theta$ is harmonic about $\Theta$ coordinator. So, generally speaking, the e velocity $u^\Theta$ is a wave. Therefore, a phase shift to show the difference for the above two sets of motion equations should be introduced.

If this physical stable condition is met, the following geometrical condition must be met.

$$\frac{\partial}{\partial t}[(\frac{R^2+r^2+2rR\cos\theta}{\cos\varphi}-1)u^\Theta]=0 \tag{65}$$

It says that: the smoke circle vortex is stable in physical sense and also in geometrical sense.

In previous discussion, this condition is assumed for static smoke circle vortex. Here, the results can be viewed as a proof.

**The above conditions show that stable smoke vortex can exist**.

To introduce a time shift, the best way is to introduce an evolution complex spatial function:

$$\frac{R^2+r^2+2rR\cos\theta}{\cos\varphi}=C(r,\theta,\Theta,t) \tag{66}$$

For physical stable smoke vortex, the complete motion equations are simplified.

In angular momentum conservation form:

$$\frac{\partial\tilde{\sigma}_r^r}{\partial r}+\frac{\partial\tilde{\sigma}_r^\theta}{\partial\theta}+(\frac{1}{r}+\frac{r+R\cos\theta}{C\cdot\cos\varphi})\tilde{\sigma}_r^r-\frac{rR\sin\theta}{C\cdot\cos\varphi}\tilde{\sigma}_r^\theta-\frac{1}{r}\tilde{\sigma}_\theta^\theta-\frac{r+R\cos\theta}{C\cdot\cos\varphi}\tilde{\sigma}_\Theta^\Theta=\rho_0\frac{\partial u^r}{\partial t} \tag{67-1}$$

$$\frac{\partial\tilde{\sigma}_\theta^r}{\partial r}+\frac{\partial\tilde{\sigma}_\theta^\theta}{\partial\theta}+\frac{r+R\cos\theta}{C\cdot\cos\varphi}\tilde{\sigma}_\theta^r-\frac{rR\sin\theta}{C\cdot\cos\varphi}\tilde{\sigma}_\theta^\theta+\frac{rR\sin\theta}{C\cdot\cos\varphi}\tilde{\sigma}_\Theta^\Theta=\rho_0 r^2\cdot\frac{\partial u^\theta}{\partial t} \tag{67-2}$$

$$\frac{\partial\tilde{\sigma}_\Theta^\Theta}{\partial\Theta}=\rho_0\frac{\partial(C\cdot u^\Theta)}{\partial t} \tag{67-3}$$

Or, in linear momentum conservation form:

$$\frac{\partial\tilde{\sigma}_r^r}{\partial r}+\frac{\partial\tilde{\sigma}_\theta^r}{\partial\theta}+(2r+R\cos\theta)\tilde{\sigma}_r^r-\frac{R}{r}\sin\theta\cdot\tilde{\sigma}_\theta^r-r\cdot\tilde{\sigma}_\theta^\theta-(r+R\cos\theta)\tilde{\sigma}_\Theta^\Theta=\rho_0\frac{\partial u^r}{\partial t} \tag{68-1}$$

$$\frac{\partial\tilde{\sigma}_r^\theta}{\partial r}+\frac{\partial\tilde{\sigma}_\theta^\theta}{\partial\theta}+(2r+R\cos\theta)\tilde{\sigma}_r^\theta+\frac{R}{r}\sin\theta\cdot(\tilde{\sigma}_\theta^\theta-\tilde{\sigma}_\Theta^\Theta)=\rho_0\frac{\partial u^\theta}{\partial t} \tag{68-2}$$

$$\frac{\partial\tilde{\sigma}_\Theta^\Theta}{\partial\Theta}=\rho_0\frac{\partial u^\Theta}{\partial t} \tag{68-3}$$

To meet all the above equations, the field quantities waiting to be determined are: $u^r$, $u^\theta$, $u^\Theta$, $\cos\varphi$, and $C$. As the number of un-known quantities is five and there are six equations, the material parameters ($\tilde{\lambda},\tilde{\mu}$) must be variables (at lest one of both is variable). The dynamic viscosity concept can be understood from this viewpoint. The smoke circle vortex will not change material features.

Based on my researches about fatigue-cracking, for solid medium, the acceptable way is to use the linear moment conservation equations to get basic solutions. Then, using the angular momentum conservation equations to get the material features variation.

If one accepts the two sets equation as equal weight, the only conclusion is that: **deformation will cause material feature variation. This is fact.**

In fact, based on previous research, the **Static Stability Condition of Smoke Vortex** is:

$$\frac{\partial}{\partial}[(\frac{R^2+r^2+2rR\cos\theta}{\cos\varphi}-1)u^\Theta]=0 \text{, and } \frac{\partial}{\partial\Theta}\left(\frac{1}{\cos\varphi}\right)=0 \tag{69}$$

By this way, the intrinsic instability is attributed to the sub-shear motion. On mathematic sense, this point is not intrinsic stable and it is temporal. In chaos dynamic research, this condition is viewed as focus of dynamic traces.

However, for deterministic studies, a deterministic featured dynamic equation is looked for. Hence, an alternative way should be found.

**4.4 Average Dynamic Motion Equations**

A reasonable way is the linear moment conservation and angular moment conservation are met in equal possibility. Hence, the solution must be formulated on average sense.

Based on this point, the average motion equation can be obtained by:

$$\tilde{\sigma}_i^j\big|_j + \tilde{\sigma}_j^i\big|_j = \frac{\rho_0}{\cos\varphi}\frac{\partial}{\partial t}(g_{jl}^0 R_i^l u^j) + \rho_0 \frac{\partial u^i}{\partial t} \tag{70-1}$$

As the shear stress is anti-symmetrical, the average stress non-zero components are:

$$\overline{\sigma}_r^r = \overline{\sigma}_\theta^\theta = \tilde{\lambda}\left(\frac{1}{\cos\varphi}-1\right), \quad \overline{\sigma}_\Theta^\Theta = (\tilde{\lambda}+2\tilde{\mu})\left(\frac{1}{\cos\varphi}-1\right) \tag{70-2}$$

The motion equations are:

$$\frac{\partial \overline{\sigma}_r^r}{\partial r} + \frac{1}{2}(2r+R\cos\theta+\frac{1}{r}+\frac{r+R\cos\theta}{R^2+r^2+2rR\cos\theta})\overline{\sigma}_r^r$$
$$-\frac{1}{2}(r+\frac{1}{r})\overline{\sigma}_\theta^\theta - \frac{1}{2}(r+R\cos\theta+\frac{r+R\cos\theta}{R^2+r^2+2rR\cos\theta})\overline{\sigma}_\Theta^\Theta = \rho_0 \frac{\partial u^r}{\partial t} \tag{71-1}$$

$$\frac{\partial \overline{\sigma}_\theta^\theta}{\partial \theta} + \frac{1}{2}(\frac{R}{r}\cos\theta - \frac{rR\sin\theta}{R^2+r^2+2rR\cos\theta})(\overline{\sigma}_\theta^\theta - \overline{\sigma}_\Theta^\Theta) = \frac{\rho_0(1+r^2)}{2}\cdot\frac{\partial u^\theta}{\partial t} \tag{71-2}$$

$$\frac{\partial \overline{\sigma}_\Theta^\Theta}{\partial \Theta} = \frac{\rho_0}{2}\frac{\partial}{\partial t}[(1+\frac{R^2+r^2+2rR\cos\theta}{\cos\varphi})\cdot u^\Theta] \tag{71-3}$$

If physical stability condition Equations (65) and (66) are used, the equations become:

$$\frac{\partial \overline{\sigma}_r^r}{\partial r} + \frac{1}{2}(2r+R\cos\theta+\frac{1}{r}+\frac{r+R\cos\theta}{C\cdot\cos\varphi})\overline{\sigma}_r^r$$
$$-\frac{1}{2}(r+\frac{1}{r})\overline{\sigma}_\theta^\theta - \frac{1}{2}(r+R\cos\theta+\frac{r+R\cos\theta}{C\cdot\cos\varphi})\overline{\sigma}_\Theta^\Theta = \rho_0 \frac{\partial u^r}{\partial t} \tag{72-1}$$

$$\frac{\partial \overline{\sigma}_\theta^\theta}{\partial \theta} + \frac{1}{2}(\frac{R}{r}\cos\theta - \frac{rR\sin\theta}{C\cdot\cos\varphi})(\overline{\sigma}_\theta^\theta - \overline{\sigma}_\Theta^\Theta) = \frac{\rho_0(1+r^2)}{2}\cdot\frac{\partial u^\theta}{\partial t} \tag{72-2}$$

$$\frac{\partial \overline{\sigma}_\Theta^\Theta}{\partial \Theta} = \rho_0 \frac{\partial}{\partial t}(\frac{1+C}{2}u^\Theta) \tag{72-3}$$

There are four unknown quantities. Hence, an additional equation should be supplied. The best way is to introduce dynamic energy conservation equation:

$$\frac{\rho_0}{2}[(u^r)^2+(ru^\theta)^2+C\cdot\cos\varphi\cdot(u^\Theta)^2] = E_0 \tag{72-4}$$

Other ways are possible also.

It shows that: for solve smoke dynamic evolution problems, the best way is to introduce an energy distribution determined evolution function $C(r,\theta,\Theta)$. In fact, many turbulence researches are based on this view-point to explain the turbulence generation process.

## 5. Dynamic Motion Equations in Viscous Fluid

For viscous fluid materials, the problem is largely simplified. In this case, $\tilde{\lambda}=0$. The non-zero stress components are:

$$\tilde{\sigma}_\Theta^\Theta = 2\tilde{\mu}\cdot\left(\frac{1}{\cos\varphi}-1\right) \tag{73-1}$$

$$\tilde{\sigma}_\theta^r = 2\tilde{\mu}\cdot\tan\varphi \tag{73-2}$$

$$\tilde{\sigma}_r^\theta = -2\tilde{\mu}\cdot\tan\varphi \tag{73-3}$$

The average dynamic motion equations are simplified as:

$$-\frac{1}{2}(r+R\cos\theta+\frac{r+R\cos\theta}{C\cdot\cos\varphi})\bar{\sigma}_\Theta^\Theta = \rho_0\frac{\partial u^r}{\partial t} \tag{74-1}$$

$$-\frac{1}{2}(\frac{R}{r}\cos\theta-\frac{rR\sin\theta}{C\cdot\cos\varphi})\bar{\sigma}_\Theta^\Theta = \frac{\rho_0(1+r^2)}{2}\cdot\frac{\partial u^\theta}{\partial t} \tag{74-2}$$

$$\frac{\partial\bar{\sigma}_\Theta^\Theta}{\partial\Theta} = \rho_0\frac{\partial}{\partial t}(\frac{1+C}{2}u^\Theta) \tag{74-3}$$

$$\frac{\rho_0}{2}[(u^r)^2+(ru^\theta)^2+C\cdot\cos\varphi\cdot(u^\Theta)^2] = E_0 \tag{74-4}$$

Using the above four equations, with suitable boundary and initial condition, the problem is closed. They form the average stress field method.

The linear moment conservation equations are simplified as:

$$\frac{\partial\tilde{\sigma}_\theta^r}{\partial\theta}-\frac{R}{r}\sin\theta\cdot\tilde{\sigma}_\theta^r-(r+R\cos\theta)\tilde{\sigma}_\Theta^\Theta = \rho_0\frac{\partial u^r}{\partial t} \tag{75-1}$$

$$\frac{\partial\tilde{\sigma}_r^\theta}{\partial r}+(2r+R\cos\theta)\tilde{\sigma}_r^\theta-\frac{R}{r}\sin\theta\cdot\tilde{\sigma}_\Theta^\Theta = \rho_0\frac{\partial u^\theta}{\partial t} \tag{75-2}$$

$$\frac{\partial\tilde{\sigma}_\Theta^\Theta}{\partial\Theta} = \rho_0\frac{\partial u^\Theta}{\partial t} \tag{75-3}$$

The angular moment conservation equations are simplified as:

$$\frac{\partial\tilde{\sigma}_r^\theta}{\partial\theta}-\frac{rR\sin\theta}{C\cdot\cos\varphi}\tilde{\sigma}_r^\theta-\frac{r+R\cos\theta}{C\cdot\cos\varphi}\tilde{\sigma}_\Theta^\Theta = \rho_0\frac{\partial u^r}{\partial t} \tag{76-1}$$

$$\frac{\partial\tilde{\sigma}_\theta^r}{\partial r}+\frac{r+R\cos\theta}{C\cdot\cos\varphi}\tilde{\sigma}_\theta^r+\frac{rR\sin\theta}{C\cdot\cos\varphi}\tilde{\sigma}_\Theta^\Theta = \rho_0 r^2\cdot\frac{\partial u^\theta}{\partial t} \tag{76-2}$$

$$\frac{\partial\tilde{\sigma}_\Theta^\Theta}{\partial\Theta} = \rho_0\cdot\frac{\partial(Cu^\Theta)}{\partial t} \tag{76-3}$$

Combing the above six equations, with suitable boundary and initial condition, the problem is closed. They form the exact stress field method.

Note that: if omitting the shear stress, replacing the $\rho_0$ with $\rho_0 C$, and replacing $\frac{\partial u^r}{\partial t}$ and $\frac{\partial u^\theta}{\partial t}$ with $\cos\varphi \cdot \frac{\partial u^r}{\partial t}$ and $\cos\varphi \cdot \frac{\partial u^\theta}{\partial t}$, they have similar mathematic structure.

This means that: the classical motion equations, such as NS equations, can be used as an approximation.

From this fact, it is easy to understand why the smoke circle vortex can be approximated by so many ways as they are shown by huge documents.

For stable static smoke vortex, through local rotation parameter $\cos\varphi$, the dynamic equations control the time evolution of shear flow, firstly. Then, the geometrical equations discussed in previous sections will determine the spatial evolution. Finally, the local rotation parameter $\cos\varphi$ is changed. In some sense, these two processes are not so directly coupled as in solid mechanics.

As the wave velocity in fluid is limited, a time lag is sure. Hence, it can expect that: shear wave velocity of medium plays a very important role in dynamic evolution of smoke vortex.

## 6. Approximation Equations near Smoke Center Circle in Viscous Fluid

Near the smoke center circle, $r \ll R$.

Under this condition,

$$\frac{R^2 + r^2 + 2rR\cos\theta}{\cos\varphi} \approx \frac{R^2}{\cos\varphi} = C(\Theta, t) \tag{77}$$

Putting the stress items into them, the related motion equations are discussed bellow. It is clear, the C has very big amplitude.

### 6.1 Angular Momentum Conservation Equations

The angular moment conservation equations are simplified as:

$$-\frac{\partial \tan\varphi}{\partial \theta} - \frac{\cos\theta}{R}(\frac{1}{\cos\varphi} - 1) = \frac{\rho_0}{2\tilde{\mu}} \frac{\partial u^r}{\partial t} \tag{78-1}$$

$$\frac{\partial \tan\varphi}{\partial r} + \frac{\cos\theta}{R}\tan\varphi = \frac{\rho_0}{2\tilde{\mu}} r^2 \cdot \frac{\partial u^\theta}{\partial t} \tag{78-2}$$

$$\frac{\partial}{\partial \Theta}(\frac{1}{\cos\varphi}) = \frac{\rho_0}{2\tilde{\mu}} \cdot \frac{\partial}{\partial t}(\frac{R^2 \cdot u^\Theta}{\cos\varphi}) \tag{78-3}$$

Note that, the smoke vortex center circle radium is time dependent. $R = R(t)$.

The first two equations give out a relation equation for shear velocity:

$$-\frac{1}{R}\{\frac{\partial}{\partial r}\cos\theta(\frac{1}{\cos\varphi} - 1)] - \frac{\partial}{\partial \theta}(\cos\theta \cdot \tan\varphi)\} = \frac{\rho_0}{2\tilde{\mu}} \frac{\partial}{\partial t}(\frac{\partial u^r}{\partial r} + r^2 \frac{\partial u^\theta}{\partial \theta}) \tag{79-1}$$

For large R, omitting the local variation of local rotation, its approximation equation is:

$$\frac{\partial u^r}{\partial r} + r^2 \frac{\partial u^\theta}{\partial \theta} = 0 \tag{79-2}$$

It is the incompressibility condition for shear flow. In fact, this equation is widely used in turbulence researches.

By perturbation view-point, observing Equation (79-1), the smoke vortex is produced by the plane compressibility variation of fluid. This view-point is very common.

The Equation (78-3) is a non-linear wave equation about $\frac{1}{\cos\varphi}$. To show this clear, it is rewritten as:

$$\frac{\partial}{\partial \Theta}(\frac{1}{\cos\varphi}) = \frac{\rho_0}{2\tilde{\mu}} \cdot R^2 u^\Theta \cdot \frac{\partial}{\partial t}(\frac{1}{\cos\varphi}) + \frac{1}{\cos\varphi}\frac{\partial}{\partial t}(\frac{\rho_0}{2\tilde{\mu}} \cdot R^2 \cdot u^\Theta) \qquad (80)$$

It shows that: the $\frac{1}{\cos\varphi}$ wave transportation speed is: $V_{tra} = \frac{2\tilde{\mu}}{R^2 u^\Theta}$, and the wave production disturbance is the variation of wave transportation speed. Using the transportation speed, the equation can be rewritten as the **general evolution equation of smoke vortex:**

$$\frac{\partial}{\partial \Theta}(\frac{1}{\cos\varphi}) = \frac{1}{V_{tra}} \cdot \frac{\partial}{\partial t}(\frac{1}{\cos\varphi}) + \frac{1}{\cos\varphi}\frac{\partial}{\partial t}(\frac{1}{V_{tra}}) \qquad (81)$$

This equation is very general form for smoke evolution. Note that it is universal.

### 6.2 Linear Momentum Conservation Equations

The linear moment conservation equations are simplified as:

$$-R\sin\theta \cdot \tan\varphi = \frac{\rho_0}{2\tilde{\mu}} r \frac{\partial u^r}{\partial t} \qquad (82\text{-}1)$$

$$-R\sin\theta \cdot (\frac{1}{\cos\varphi} - 1) = \frac{\rho_0}{2\tilde{\mu}} r \frac{\partial u^\theta}{\partial t} \qquad (82\text{-}2)$$

$$\frac{\partial}{\partial \Theta}(\frac{1}{\cos\varphi}) = \frac{\rho_0}{2\tilde{\mu}} \frac{\partial u^\Theta}{\partial t} \qquad (82\text{-}3)$$

The first two equations give out a relation equation for shear velocity:

$$\tan\varphi \cdot \frac{\partial u^\theta}{\partial t} = (\frac{1}{\cos\varphi} - 1)\frac{\partial u^r}{\partial t} \qquad (84)$$

Comparing Equation (82-3) and Equation (80), if both linear momentum conservation and angular momentum conservation be met, the necessary condition is:

$$(1 - \frac{R^2}{\cos\varphi})\frac{\partial u^\Theta}{\partial t} = u^\Theta \cdot \frac{\partial}{\partial t}(1 - \frac{R^2}{\cos\varphi}) \qquad (85\text{-}1)$$

Or, in the form:

$$\frac{\partial}{\partial t}\left(\frac{\cos\varphi \cdot u^\Theta}{\cos\varphi - R^2}\right) = 0 \qquad (85\text{-}2)$$

For large R, it is approximated as:

$$\frac{\partial}{\partial t}\left(\frac{\cos\varphi \cdot u^\Theta}{R^2}\right) = 0 \qquad (85\text{-}3)$$

It gives out a **general circling velocity conservation rule**:

$$u^\Theta = \frac{R^2(t)}{\cos\varphi} \cdot \left(\frac{\cos\varphi(0)}{R^2(0)} u^\Theta(0)\right) \qquad (85\text{-}4)$$

Where, the zero in brackets shows the initial value for simplicity. The circling velocity is completely determined by initial condition.

### 6.3 Dynamic Evolution Equations for Exact Stress Field Methods

Summering the results of above sub-sections, the dynamic evolution equations can be obtained. For the linear momentum and angular momentum equations should be satisfied at the same time, **the dynamic evolution equations for exact stress field methods** is obtained as:

$$\frac{\partial}{\partial t}\left(\frac{\cos\varphi \cdot u^\Theta}{R^2}\right) = 0 \qquad (86\text{-}1)$$

$$\frac{\partial}{\partial \Theta}(\frac{1}{\cos \varphi}) = \frac{\rho_0}{2\tilde{\mu}} \cdot \frac{\partial}{\partial t}(\frac{R^2 \cdot u^\Theta}{\cos \varphi}) \quad (86\text{-}2)$$

$$\frac{\partial u^r}{\partial r} + r^2 \frac{\partial u^\theta}{\partial \theta} = 0 \quad (86\text{-}3)$$

$$\tan \varphi \cdot \frac{\partial u^\theta}{\partial t} = (\frac{1}{\cos \varphi} - 1)\frac{\partial u^r}{\partial t} \quad (86\text{-}4)$$

Where, $R = R(t)$ is the smoke circle radium evolution about time. As the condition $R \gg r$ is assumed, the $\frac{1}{\cos \varphi} = \frac{1}{\cos \varphi(\Theta, t)}$ is only the function of global coordinator $\Theta$ and time $t$. The motion on circle is independent from the disturbed motion around the circle. Once the circle solution is obtained, the $\varphi(\Theta, t)$ will determine the subscale motion around the center circle.

Repeating this method, a series of sub-scale cascade motion equations can be established. Then, multi-scale smoke vortexes problems can be formulated.

As the average stress method is easy to be performed based on statistic data, the following section will give out its equations.

**6.4 Dynamic Evolution Equations for Average Stress Field Methods**

Near the smoke vortex center, the average stress dynamic motion equations are simplified as:

$$-(R\cos\theta + \frac{\cos\theta}{R})(\frac{1}{\cos\varphi} - 1) = \frac{\rho_0}{\tilde{\mu}} \frac{\partial u^r}{\partial t} \quad (87\text{-}1)$$

$$-R\cos\theta \cdot (\frac{1}{\cos\varphi} - 1) = \frac{\rho_0 r(1+r^2)}{2\tilde{\mu}} \cdot \frac{\partial u^\theta}{\partial t} \quad (87\text{-}2)$$

$$\frac{\partial}{\partial \Theta}(\frac{1}{\cos\varphi}) = \frac{\rho_0}{4\tilde{\mu}} R^2 \frac{\partial}{\partial t}[(1 + \frac{1}{\cos\varphi})u^\Theta] \quad (87\text{-}3)$$

$$\frac{\rho_0}{2}[(u^r)^2 + (ru^\theta)^2 + R^2 \cdot (u^\Theta)^2] = E_0 \quad (87\text{-}4)$$

For very large R ($R \gg 1$), the first two equations give out relation equation:

$$r(1+r^2)\frac{\partial u^\Theta}{\partial t} = 2\frac{\partial u^r}{\partial t} \quad (88)$$

Therefore, the **dynamic evolution equations for average stress field method** are:

$$\frac{\partial}{\partial \Theta}(\frac{1}{\cos\varphi}) = \frac{\rho_0}{4\tilde{\mu}} \frac{\partial}{\partial t}[(1 + \frac{R^2}{\cos\varphi})u^\Theta] \quad (89\text{-}1)$$

$$\frac{\rho_0}{2}[(u^r)^2 + (ru^\theta)^2 + R^2 \cdot (u^\Theta)^2] = E_0 \quad (89\text{-}2)$$

$$r(1+r^2)\frac{\partial u^\Theta}{\partial t} = 2\frac{\partial u^r}{\partial t} \quad (89\text{-}3)$$

If omitting the sub-shear energy and motion (view them as noise), they can be further simplified as:

$$\frac{\partial}{\partial \Theta}(\frac{1}{\cos\varphi}) = \frac{\rho_0}{4\tilde{\mu}} \frac{\partial}{\partial t}[(1 + \frac{R^2}{\cos\varphi})u^\Theta] \quad (90\text{-}1)$$

$$\frac{\rho_0}{2} R^2 \cdot (u^\Theta)^2 = E_0 \quad (90\text{-}2)$$

This is the **simplest approximation equations for smoke vortex circle center evolution**. However, its non-linear feature is significant. The last equation says that:

$\frac{\rho_0}{2} R^2 (u^\Theta)^2 = \frac{\rho_0}{2}(V^\Theta)^2 = E_0$, the macro displacement velocity amplitude is constant along the circulation direction. As the $u^\Theta$ is angular velocity, the Equation (90-2) shows: the angular velocity is decreased for large radium circle in square-inverse laws. Hence, for small circle, the circle rotates very fast. This picture forms a strong contrast with the constant macro displacement velocity amplitude.

### 7. Conclusion

The general dynamic evolution equations for smoke vortex are formulated. They are dynamic evolution equations for exact stress field method and dynamic evolution equations for average stress field method. For industrial application and experimental data processing, their corresponding approximation equations for viscous fluid are given.

As the researches on this topic is huge, I apologies for not giving out detailed reference papers. I hope this paper will promote further researches on smoke circle vortex and turbulence.